\documentclass[a4paper,11pt]{article}
\usepackage{pos}
\usepackage{subfigure}
\usepackage{soul}

\title{
    Tensor renormalization group approach to the $O(2)$ models via symmetry-twisted partition functions
}
\ShortTitle{TRG for the $O(2)$ models via symmetry-twisted partition functions}

\usepackage{tikz,xcolor}
\definecolor{lime}{HTML}{A6CE39}
\DeclareRobustCommand{\orcidicon}{%
	\begin{tikzpicture}
	\draw[lime, fill=lime] (0,0) 
	circle [radius=0.16] 
	node[white] {{\fontfamily{qag}\selectfont \tiny ID}};	\draw[white, fill=white] (-0.0625,0.095) 
	circle [radius=0.007];	\end{tikzpicture}
	\hspace{-2mm}}
\foreach \x in {A,B,C,D,E}{%
	\expandafter\xdef\csname orcid\x\endcsname{\noexpand\href{https://orcid.org/\csname orcidauthor\x\endcsname}{\noexpand\orcidicon}}
}

\newcommand{\FNAL}{\affiliation[g]{
Fermi National Accelerator Laboratory, Batavia, Illinois, USA}}

\newcommand{\Cone}{\affiliation[f]{Capital One, Chicago, Illinois, USA}}

\newcommand{\NCSU}{\affiliation[c]{
Department of Physics and Astronomy, North Carolina State University, Raleigh, North Carolina 27695, USA}}

\newcommand{\TU}{\affiliation[a]{Center for Computational Sciences, University of Tsukuba, Tsukuba, Ibaraki 305-8577, Japan}}

\newcommand{\UTokyo}{\affiliation[b]{Graduate School of Science, The University of Tokyo, Bunkyo-ku, Tokyo, 113-0033, Japan}}

\newcommand{\KU}{\affiliation[d]{Department of Physics,
Kyoto University, Kyoto, 606-8502, Japan}}

\author*[a,b]{{Shinichiro Akiyama}~\orcidA{}}
\emailAdd{akiyama@ccs.tsukuba.ac.jp}
\TU
\UTokyo

\author[c]{Raghav G. Jha~\orcidB{}}
\emailAdd{raghav.govind.jha@gmail.com}
\NCSU

\author[d]{Jun Maeda~\orcidC{}}
\emailAdd{maeda@gauge.scphys.kyoto-u.ac.jp}
\KU

\author[e]{Yuya Tanizaki~\orcidD{}}
\emailAdd{yuya.tanizaki@yukawa.kyoto-u.ac.jp}
\affiliation[e]{Yukawa Institute for Theoretical Physics, Kyoto University, Kyoto 606-8502, Japan}

\author[f,g]{Judah Unmuth-Yockey~\orcidE{}}
\emailAdd{jfunmuthyockey@gmail.com}
\Cone\FNAL

\notes{
    \note{
        UTHEP-819, UTCCS-P-176, KUNS-3093
    }
}

\abstract{
    We investigate critical phenomena in the $O(2)$ models using symmetry-twisted partition functions that can be efficiently computed within the tensor renormalization group framework.
    We first demonstrate, taking the three-dimensional model as an example, that symmetry-twisted partition functions detect the spontaneous breaking of global continuous symmetry.
    We then consider the same model in two dimensions, where the Berezinskii--Kosterlitz--Thouless (BKT) transition occurs.
    Since symmetry-twisted partition functions directly provide the helicity modulus at a finite twist angle, we determine the BKT transition point. 
    These results are presented based on Ref.~\cite{Akiyama:2026dzg}.
    Finally, in addition to the original paper~\cite{Akiyama:2026dzg}, we apply this approach to the two-dimensional generalized $O(2)$ model and confirm that it successfully identifies the phase transitions between the ferromagnetic and nematic phases, as well as between the nematic and paramagnetic phases.
}

\FullConference{The 42nd International Symposium on Lattice Field Theory (LATTICE2025)\\
2-8 November 2025\\
Tata Institute of Fundamental Research, Mumbai, India\\}

\begin{document}
\maketitle

\section{Introduction}

Tensor networks provide a numerical real-space renormalization-group approach to investigating lattice field theories, serving as an alternative to Monte Carlo (MC) methods.
A striking advantage of tensor networks is that they are useful even in the presence of the sign problem and for complex actions, which are encountered in many interesting problems in lattice QCD.
By formulating quantum many-body systems in terms of tensor network representations, the range of observables and quantities that can be efficiently accessed differs significantly from those accessible by MC methods.
In particular, tensor networks allow direct access to partition functions~\cite{Levin:2006jai,Adachi:2019paf,Adachi:2020upk}, which are generally difficult to evaluate within the MC framework.
This advantage was utilized to calculate the ground-state degeneracy solely from ratios of partition functions~\cite{Gu:2009dr}: When a discrete global symmetry $G$ is spontaneously broken to a subgroup $H$, 
\begin{align}
\label{eq:Gu-Wen}
    \frac{Z(L_{1},\cdots,L_{d})^{2}}{Z(L_{1},\cdots,2L_{d})} = |G/H|.
\end{align}
follows in the thermodynamic limit, where $Z(L_{1},\cdots,L_{d})$ denotes the partition function on a torus $T^{d}$ with the size $L_{1}\times L_{2}\times\cdots\times L_{d}$.
We refer to the ratio in Eq.~\eqref{eq:Gu-Wen} as the Gu--Wen ratio, which counts the number of degenerate vacua.
This Gu--Wen ratio is practically useful for locating the critical point associated with the spontaneous breaking of discrete symmetries.
Moreover, finite-size scaling of the Gu--Wen ratio enables one to estimate not only the critical point but also the shift exponent~\cite{Morita:2024lwg}. 
Its universal value can be obtained from conformal field theory (CFT) via modular-invariant partition functions~\cite{Morita:2025hsv}.

The Gu--Wen ratio is not the unique way to extract the low-energy symmetry realization, and the symmetry-twisted partition functions also provide the useful computational framework for the same purpose.
Suppose we are interested in a local field theory with global symmetries. We can then introduce a background gauge field $A$ and denote by $Z[A]$ the partition function in the presence of this background.
Let us consider the same situation as in Eq.~\eqref{eq:Gu-Wen}, where each vacuum is specified by an element of the left coset $G/H=\{gH|g\in G\}$.
Hereafter, we focus on flat background gauge fields $A$, characterized by the holonomies $(g_{1}, g_{2}, \cdots, g_{d})$ along each cycle of $T^{d}$. 
By choosing $A=(1, \cdots, 1, g_{0})$ with $g_{0}\in G$, a symmetry twist is introduced only along the $d$-direction.
The symmetry-twisted partition function $Z_{g_0}=Z[A=(1, \cdots, 1, g_{0})]$ in the thermodynamic limit evaluates
\begin{align}
\label{eq:Zg0}
    Z_{g_0}
    =
    |\{gH\in G/H|g^{-1}g_{0}g\in H\}|
    \exp\left[
        -\int_{T^{d}}\Lambda~{\rm d}^{d}x
    \right],
\end{align}
where the prefactor gives the universal low-energy data that counts the number of vacua on the symmetry-twisted torus without any domain wall.
In contrast, the energy density of the ground state $\Lambda$ is non-universal.
\footnote{
At this stage, the Gu--Wen ratio can be regarded as one of the simplest ratios, $Z_{1}(L_{1},\cdots,L_{d})^{2}/Z_{1}(L_{1},\cdots,2L_{d})$, that eliminates such non-universal contributions.
}
Therefore, 
\begin{align}
\label{eq:ST_Z}
    \frac{Z_{g_{0}}(L_{1},\cdots,L_{d})}{Z_{1}(L_{1},\cdots,L_{d})}
    =
    \frac{|\{gH\in G/H|g^{-1}g_{0}g\in H\}|}{|G/H|}
\end{align}
holds in the thermodynamic limit.
When $g^{-1}Hg=H$, Eq.~\eqref{eq:ST_Z} further simplifies to 
\begin{align}
\label{eq:jump}
    \frac{Z_{g_{0}}}{Z_{1}}
    =
    \begin{cases}
        1, ~~~ g_{0}\in H,\\
        0, ~~~ g_{0}\notin H.
    \end{cases}
\end{align}
In the symmetric phase, $Z_{g_{0}}/Z_{1} = 1$ for any $g_{0} \in G$, while it vanishes in the symmetry-broken phase. The resulting discontinuity marks the phase transition.
As demonstrated in Refs.~\cite{Maeda:2025ycr,Akiyama:2026dzg}, the same conclusion as in Eq.~\eqref{eq:jump} applies when a global U(1) symmetry is spontaneously broken.
Ref.~\cite{Akiyama:2026dzg} showed that Eq.~\eqref{eq:ST_Z} can be straightforwardly computed within the tensor renormalization group (TRG) framework once the global symmetry is imposed on the fundamental tensors of the network representation of $Z_{g_{0}}$. 
This is because TRG naturally accommodates periodic boundary conditions.

\section{Models and methods}

Here, we apply symmetry-twisted partition functions to study phase transitions in the classical $O(2)$ model, whose Hamiltonian is given by
\begin{align}
\label{eq:o2}
    H=-\sum_{n\in\Gamma_{d}}\sum_{\mu=1}^{d}
    \cos(\theta_{n+\hat{\mu}}-\theta_{n}),
\end{align}
where $\theta_n$ denote U(1) variables, and $n$ labels the lattice sites of a $d$-dimensional cubic lattice $\Gamma_d$, with unit vectors $\hat{\mu}$ in the $\mu$-directions. 
In three dimensions, this model provides a prototypical example of spontaneous breaking of a continuous symmetry at finite temperature $T$.
In two dimensions, by contrast, the Berezinskii--Kosterlitz--Thouless (BKT) transition occurs without spontaneous breaking of the $O(2)$ symmetry. 
Even in this case, the symmetry-twisted partition function serves as an order parameter, since Eq.~\eqref{eq:ST_Z} yields the free-energy difference between systems with and without the symmetry twist, which closely follows the original definition of the helicity modulus, i.e., the superfluid density~\cite{Fisher1973}.

Furthermore, we consider the following generalized $O(2)$ model:
\begin{align}
\label{eq:go2}
    H=
    -\Delta\sum_{n,\mu}
    \cos(\theta_{n+\hat{\mu}}-\theta_{n})
    -(1-\Delta)\sum_{n,\mu}
    \cos\left(q(\theta_{n+\hat{\mu}}-\theta_{n})\right),
\end{align}
where $0\le\Delta\le1$ and $q$ is a positive integer.
The first term represents the standard ferromagnetic interaction, while the second, nematic term favors skew alignments with relative angles $2\pi k/q$, where $k\le q$ is an integer.
Regardless of the value of $\Delta$, the Hamiltonian in Eq.~\eqref{eq:go2} possesses the same symmetry as in Eq.~\eqref{eq:o2}.
For $\Delta=1$, the standard $O(2)$ model is recovered, whereas for $\Delta=0$, Eq.~\eqref{eq:go2} reduces to a purely nematic Hamiltonian, where the discrete rotation symmetry $\theta_{n}\mapsto\theta_{n}+2\pi/q$ is promoted to the local gauge redundancy and it cannot be spontaneously broken due to Elitzur's theorem.

For both models defined in Eqs.~\eqref{eq:ST_Z} and \eqref{eq:go2}, the corresponding tensor network representations of the twisted partition function $Z_{g_{0}}$ take the following form:
\begin{align}
\label{eq:tn}
    Z_{g_{0}}
    \simeq
    {\rm tTr}_{g_{0}}
    \left[
        \prod_{n\in\Gamma_{d}}T_{n}
    \right].
\end{align}
On the right-hand side, the tensors $T_{n}$ are defined at each lattice site $n$ with rank $2d$, and the network $\prod_{n\in\Gamma_{d}}T_{n}$ inherits the geometry of the original lattice $\Gamma_{d}$.
The symbol ${\rm tTr}_{g_{0}}$ denotes the tensor trace with the twist specified by $A = (1,\cdots,1,g_{0})$, where $g_{0} \in U(1)$.
In this study, we construct the fundamental tensor following Refs.~\cite{Liu:2013nsa,Samlodia:2024kyi}, where the truncated character expansion is employed to make the bond dimension finite, thereby allowing the global symmetry to be imposed directly on $T_n$ both for Eqs.~\eqref{eq:o2} and \eqref{eq:go2}.

The tensor contractions in Eq.~\eqref{eq:tn} are carried out approximately using TRG. To implement the twist in Eq.~\eqref{eq:tn}, we extend the TRG algorithms by incorporating symmetry blocking~\cite{Yang:2015rra,Akiyama:2026dzg}.
For $d=3$ and $d=2$, we employ the anisotropic TRG (ATRG)~\cite{Adachi:2019paf} and bond-weighted TRG (BTRG)~\cite{Adachi:2020upk} algorithms, respectively.
By parametrizing $g_{0}={\rm e}^{{\rm i}\alpha}$, the symmetry twist associated with $g_{0}$ along the $d$-direction amounts to imposing a twisted boundary condition with twist angle $\alpha$.
In what follows, we denote $Z_{g_{0}}$ by $Z_{\alpha}$ for notational simplicity.

\subsection{Three-dimensional $O(2)$ model}
\label{subsec:3d_o2}
\begin{figure}
    \centering
    \subfigure[]{
        \includegraphics[width=7.3cm]{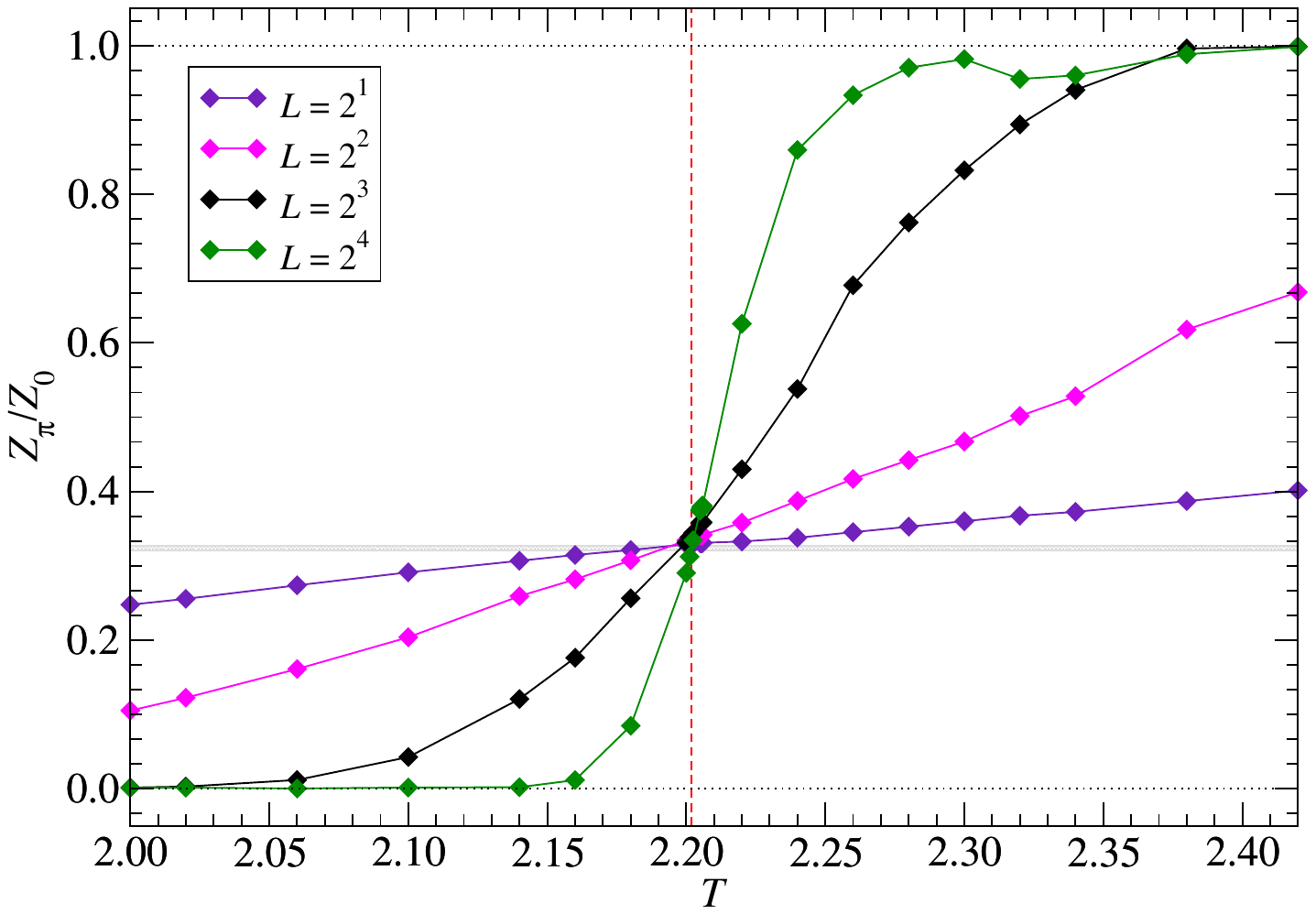}
    }
    \subfigure[]{
        \includegraphics[width=7.3cm]{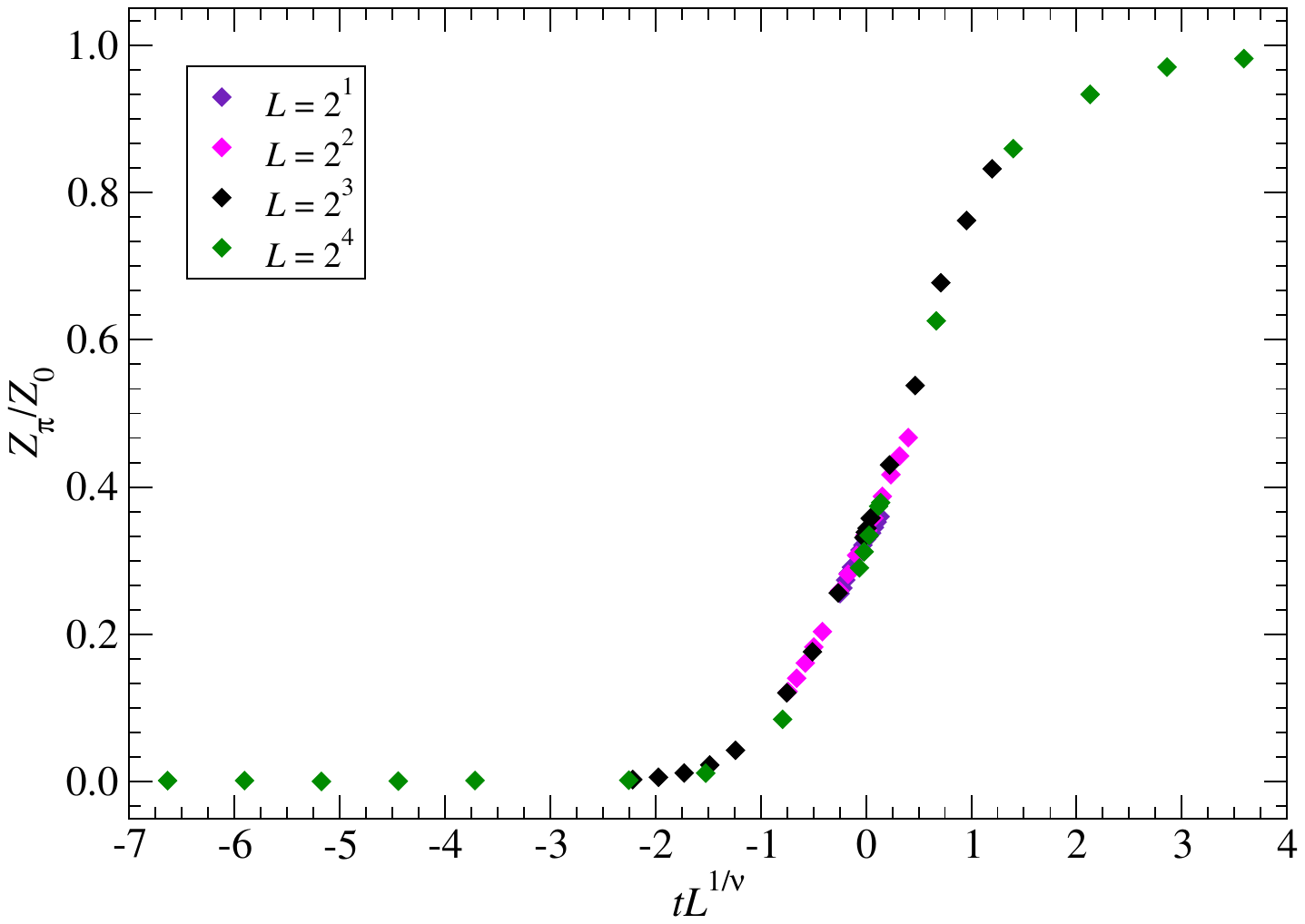}
    }
    \caption{
        ATRG results for the three-dimensional $O(2)$ model with bond dimension $D_{\rm ATRG}=96$.
        (a) $Z_{\pi}/Z_{0}$ against temperature, varying $L$.
        The gray band indicates the MC estimate of $Z_{\pi}/Z_{0}$ at criticality for $L=2^4$ from Ref.~\cite{Gottlob_1994}. 
        The red vertical dashed line denotes the estimate $T_c=2.2018441(5)$ obtained in a recent MC study~\cite{Xu:2019mvy}.
        (b) Finite-size scaling for $Z_{\pi}/Z_{0}$.
    }
    \label{fig:3d_o2} 
\end{figure}

We first consider the three-dimensional $O(2)$ model defined in Eq.~\eqref{eq:o2}. In three dimensions, the global U(1) symmetry is spontaneously broken to the trivial group at a critical temperature $T_c$.
Taking $\mathbb{Z}_{2}\subset~$U(1), we consider the twisted partition function with the twist angle $\alpha=\pi$.
Figure~\ref{fig:3d_o2}(a) shows the resulting ratio $Z_{\pi}/Z_{0}$ for several system sizes with $L_1=L_2=L_3=L$. We observe the emergence of a scale-invariant point, which signals the critical temperature $T_{c}$.
The gray band in Fig.~\ref{fig:3d_o2}(a) indicates the MC estimate of $Z_{\pi}/Z_{0}$ at criticality for $L=2^4$ from Ref.~\cite{Gottlob_1994}, which is consistent with our TRG results. The red vertical dashed line denotes the estimate $T_c=2.2018441(5)$ obtained in a recent MC study~\cite{Xu:2019mvy}.

To determine $T_{c}$, we first bracket it as $2.2012<T_{c}<2.2050$ by examining the $L$-dependence of $Z_{\pi}/Z_{0}$ in Fig.~\ref{fig:3d_o2}(a).
For a more precise estimate, we perform a finite-size scaling analysis of $Z_{\pi/Z_0}$, following the procedure described in Ref.~\cite{Akiyama:2026dzg}. 
Introducing the reduced temperature $t=(T-T_{c})/T_{c}$, the data are expected to collapse onto a universal curve when plotted against $tL^{1/\nu}$.
Scanning $T_{c}$ within the above range with step $\Delta T=0.0001$, we determine the optimal $\nu$ at each $T$ by minimizing the deviation from a universal scaling curve, following Ref.~\cite{doi:10.1143/JPSJ.62.435}. 
Since the TRG data have no statistical errors, the cost function directly measures the deviation from scaling.
The minimum deviation, $\sim6.7\times10^{-3}$, is obtained for $2.2015 \le T_c \le 2.2019$, leading to the estimates $T_{c}=2.2017(2)$ and $\nu=0.663(33)$. 
The corresponding scaling collapse is shown in Fig.~\ref{fig:3d_o2}(b), where the uncertainty is determined by allowing a $10\%$ increase of the cost function from its minimum. 
The variation of $\nu$ within the above $T_{c}$ range is negligible compared to this uncertainty.
To our knowledge, this provides the first precise TRG determination of the critical exponent $\nu$ for the three-dimensional $O(2)$ model.

\subsection{Two-dimensional $O(2)$ model}
\label{subsec:2d_o2}
Next, we consider the same $O(2)$ model in two dimensions, where the global U(1) symmetry cannot be spontaneously broken. Instead, the BKT phase with quasi-long-range order appears at low temperature.
A key advantage of our twisted partition function is that it directly provides access to the helicity modulus. 
In particular, the ratio $Z_{\alpha}/Z_{0}$ yields the helicity modulus $\Upsilon_{\alpha}$ as
\begin{align}
\label{eq:Y}
    \Upsilon_{\alpha}(L_{1},L_{2})
    =
    -T\frac{2}{\alpha^{2}}\frac{L_{2}}{L_{1}}
    \log\frac{Z_{\alpha}(L_{1},L_{2})}{Z_{0}(L_{1},L_{2})},
\end{align}
where the twist is imposed along the 2-direction, while no twist is applied along the 1-direction.
In the high-temperature region, $\Upsilon_{\alpha}$ goes to zero in the thermodynamic limit.
In contrast, in the BKT phase, $\Upsilon_{\alpha}$ remains finite, and $\Upsilon_{\alpha}(L_{1},L_{2})$ can be evaluated analytically as
\begin{align}
\label{eq:helicity_modulus}
    \Upsilon_{\alpha}(L_{1},L_{2})
    =
    \frac{T}{2\pi R^{2}}+O({\rm e}^{-\# L_{1}/L_{2}}),
\end{align}
for $0<\alpha<\pi$, where $R$ denotes the renormalized compactification radius of the boson.
Eq.~\eqref{eq:helicity_modulus} shows that for $0<\alpha<\pi$, the helicity modulus $\Upsilon_\alpha$ directly determines the renormalized compact-boson radius $R$, provided that the higher-winding contributions of $O({\rm e}^{-\# L_1/L_2})$ can be neglected. 
This approximation is justified when $L_1/L_2\gg 1$.
This relation is crucial for applying the Nelson-–Kosterlitz (NK) criterion~\cite{Nelson:1977zz} to determine the BKT transition point $T_{\rm{BKT}}$, since it amounts to comparing $\Upsilon_\alpha$ with $2T/\pi$ to identify the marginal point $R^{2}=1/4$.

\begin{figure}
    \centering
    \subfigure[]{
        \includegraphics[width=7.3cm]{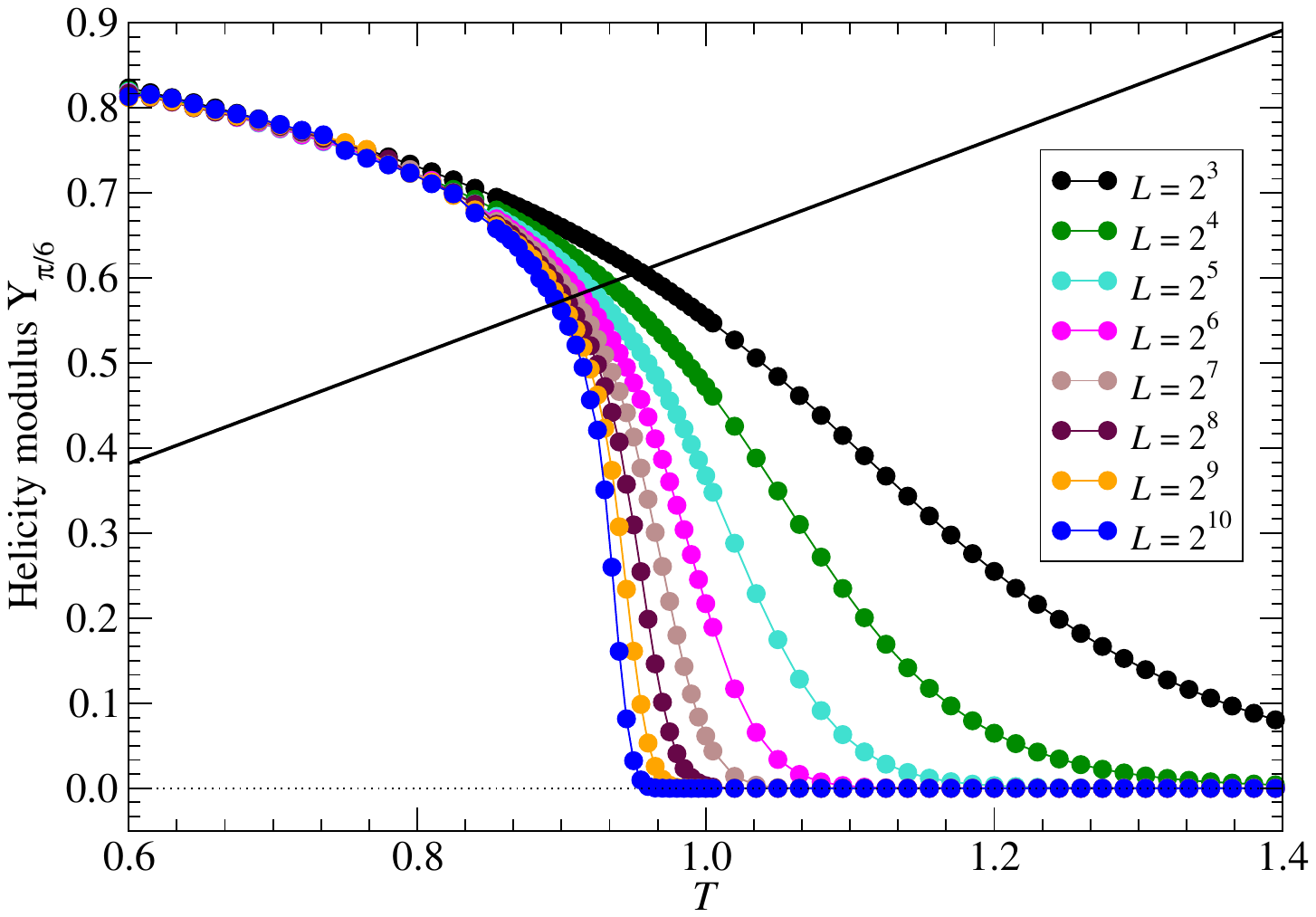}
    }
    \subfigure[]{
        \includegraphics[width=7.3cm]{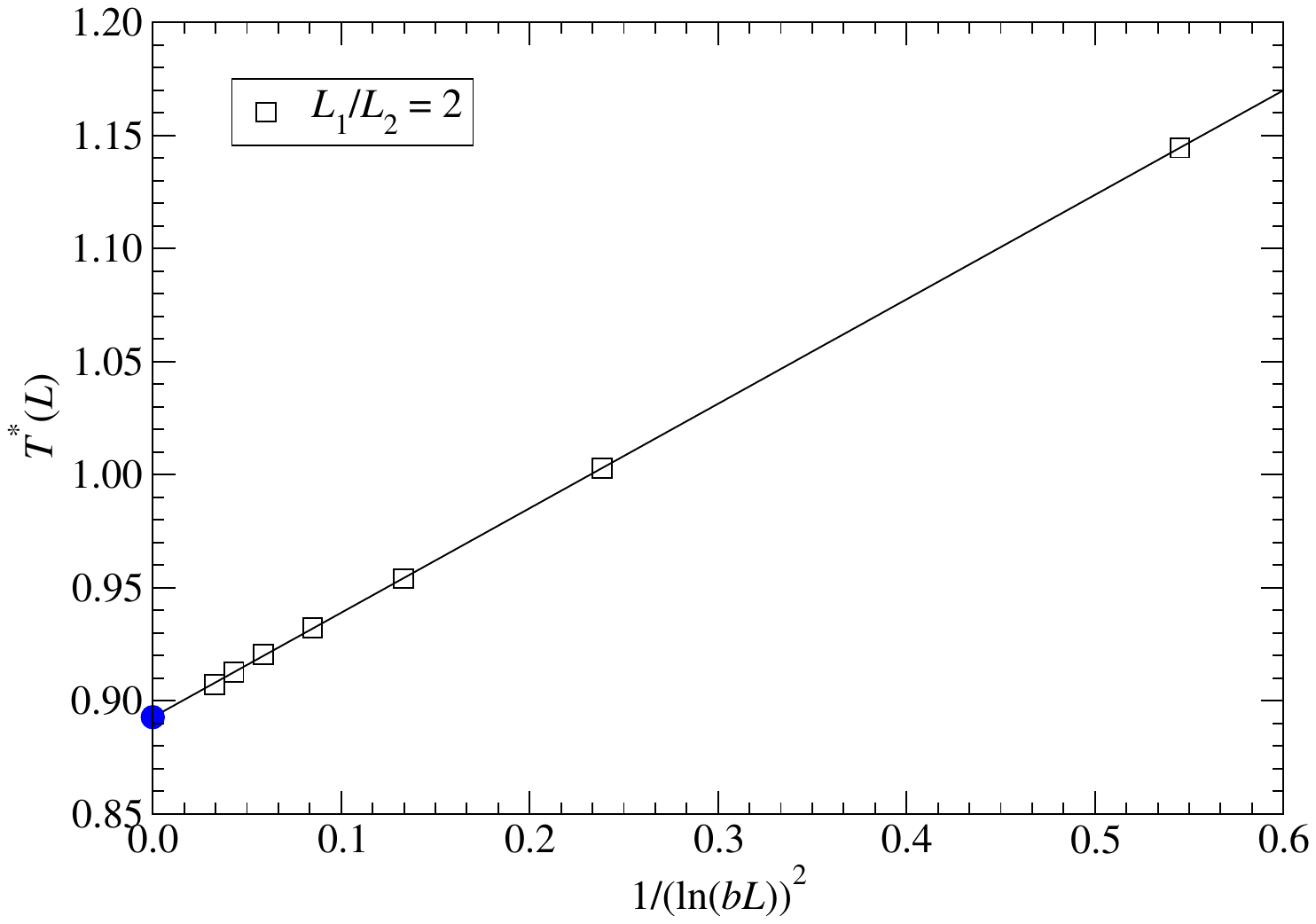}
    }
    \caption{
        BTRG results for the two-dimensional $O(2)$ model with bond dimension $D_{\rm BTRG}=200$.
        (a) $\Upsilon_{\pi/6}(2L,L)$ against temperature, varying $L$.
        The solid line shows the NK criterion, $\Upsilon_{\alpha}=2T/\pi$.
        (b) $T^{*}(L)$ as a function of $1/(\ln bL)^{2}$. 
        Squares represent $T^{*}(L)$, and the circle denotes the extrapolated $T_{\rm BKT}$ at $L\to\infty$.
    }
    \label{fig:2d_o2} 
\end{figure}

Fig.~\ref{fig:2d_o2}(a) shows the helicity modulus at $\alpha=\pi/6$ for lattices with $L_{1}=2L$ and $L_{2}=L$.
We clearly observe the ``universal jump'' originally predicted in Ref.~\cite{Nelson:1977zz}.
To estimate $T_{\rm BKT}$, we first determine the crossing temperature $T^{*}(L)$ at which $\Upsilon_{\pi/6}(2L,L)$ intersects with the NK formula.
We then extrapolate these temperatures to the thermodynamic limit $L\to\infty$, assuming the form,
\begin{align}
    T^{*}(L)=
    T_{\rm{BKT}}
    +
    \frac{a}{(\ln bL)^{2}}.
\end{align}
Fig.~\ref{fig:2d_o2}(b) shows $T^{*}(L)$ as a function of $1/(\ln bL)^{2}$.
The extrapolation to the $L\to\infty$ limit yields $T_{\rm{BKT}}= 0.8927(5)$, in good agreement with previous studies.
We remark that the dependence of $\Upsilon_{\alpha}$ on the twist angle $\alpha$ is sufficiently weak for $\alpha<\pi/4$.
We further note that $\Upsilon_{\pi}$ acquires an additional contribution, since there are two equal paths connecting antipodal points.
Taking this into account, one can determine $T_{\rm BKT}$ from $\Upsilon_{\pi}$ in a similar manner.
See Ref.~\cite{Akiyama:2026dzg} for details.

\subsection{Two-dimensional generalized $O(2)$ model}

Finally, we consider the generalized $O(2)$ model defined in Eq.~\eqref{eq:go2} with $q=2$ in two dimensions.
The phase diagram in the $\Delta$-$T$ plane exhibits three distinct phases: a paramagnetic phase, a conventional ferromagnetic phase, and a nematic phase.
In the small-$\Delta$ region, two phase transitions occur at fixed $\Delta$: from the paramagnetic to the nematic phase, and from the nematic to the ferromagnetic phase.
The former belongs to the BKT universality class, while the latter belongs to the Ising universality class.
By imposing a twist angle $\alpha\in\pi \mathbb{Z}$ or $\alpha\notin\pi\mathbb{Z}$, both phase transitions can be identified solely through the symmetry-twisted partition functions.

\begin{figure}
    \centering
    \includegraphics[width=14cm]{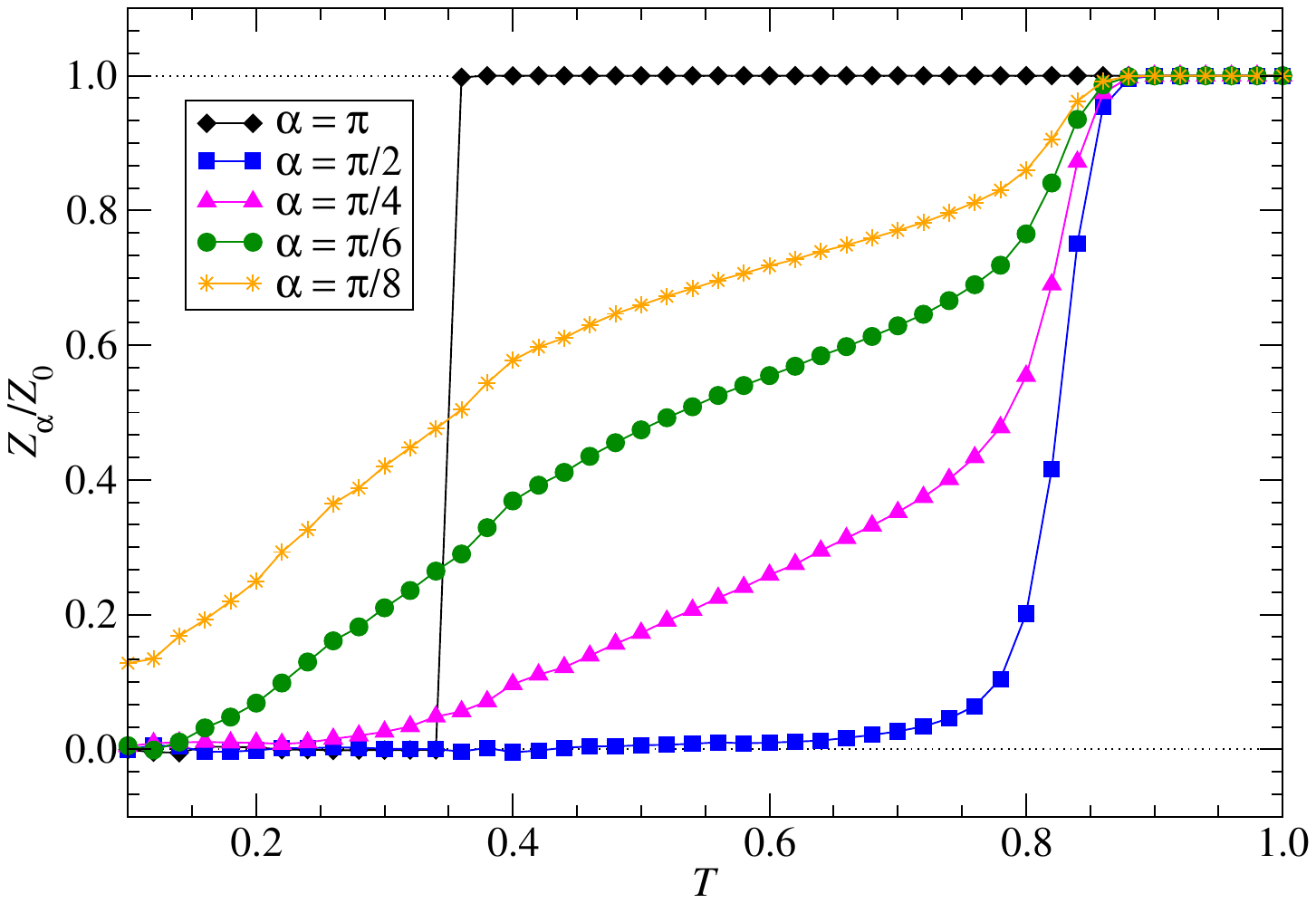}
    \caption{
        $Z_{\alpha}/Z_{0}$ at $L=2^{7}$ for the two-dimensional generalized $O(2)$ model ($q=2$, $\Delta=0.16$) with bond dimension $D_{\rm BTRG}=100$.
        Different symbols represent different twist angles $\alpha$.
    }
    \label{fig:2d_go2_zpz} 
\end{figure}

Fig.~\ref{fig:2d_go2_zpz} shows $Z_{\alpha}/Z_{0}$ at $\Delta=0.16$ for $\alpha=\pi, \pi/2, \pi/4, \pi/6,$ and $\pi/8$.
A clear jump in $Z_{\pi}/Z_{0}$ from 0 to 1 around $T\sim0.35$ signals the spontaneous breaking of the $\mathbb{Z}_{2}$ symmetry, whereas the smooth variation of $Z_{\alpha\neq \pi}/Z_{0}$ from 0 to 1 at higher temperature indicates the BKT transition.
These results demonstrate that symmetry-twisted partition functions can successfully identify the phase transitions in the generalized $O(2)$ model by appropriately choosing the twist angle $\alpha\in\pi\mathbb{Z}$ or $\alpha\notin\pi\mathbb{Z}$.

\begin{figure}
    \centering
    \subfigure[]{
    \includegraphics[width=7.3cm]{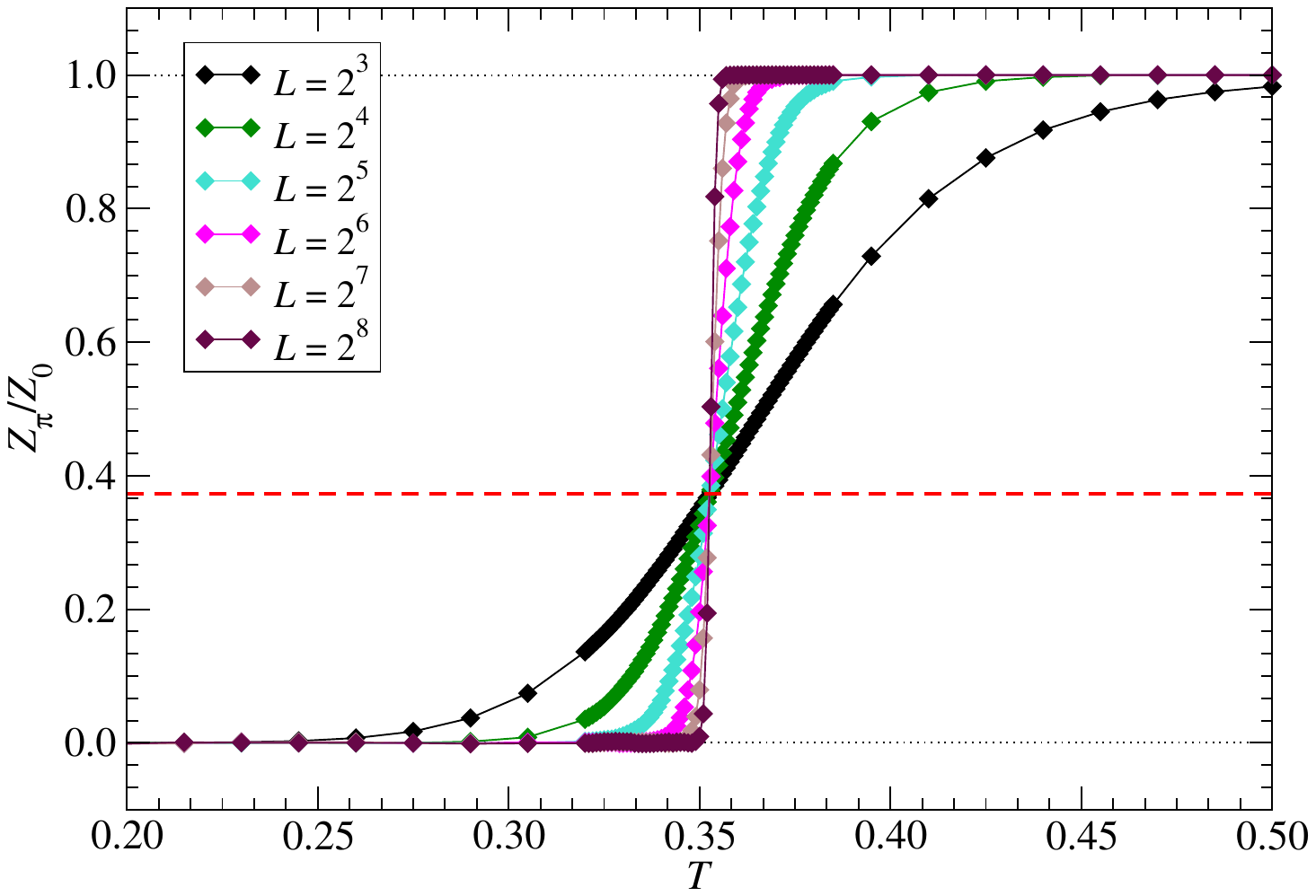}
    }
    \subfigure[]{
    \includegraphics[width=7.3cm]{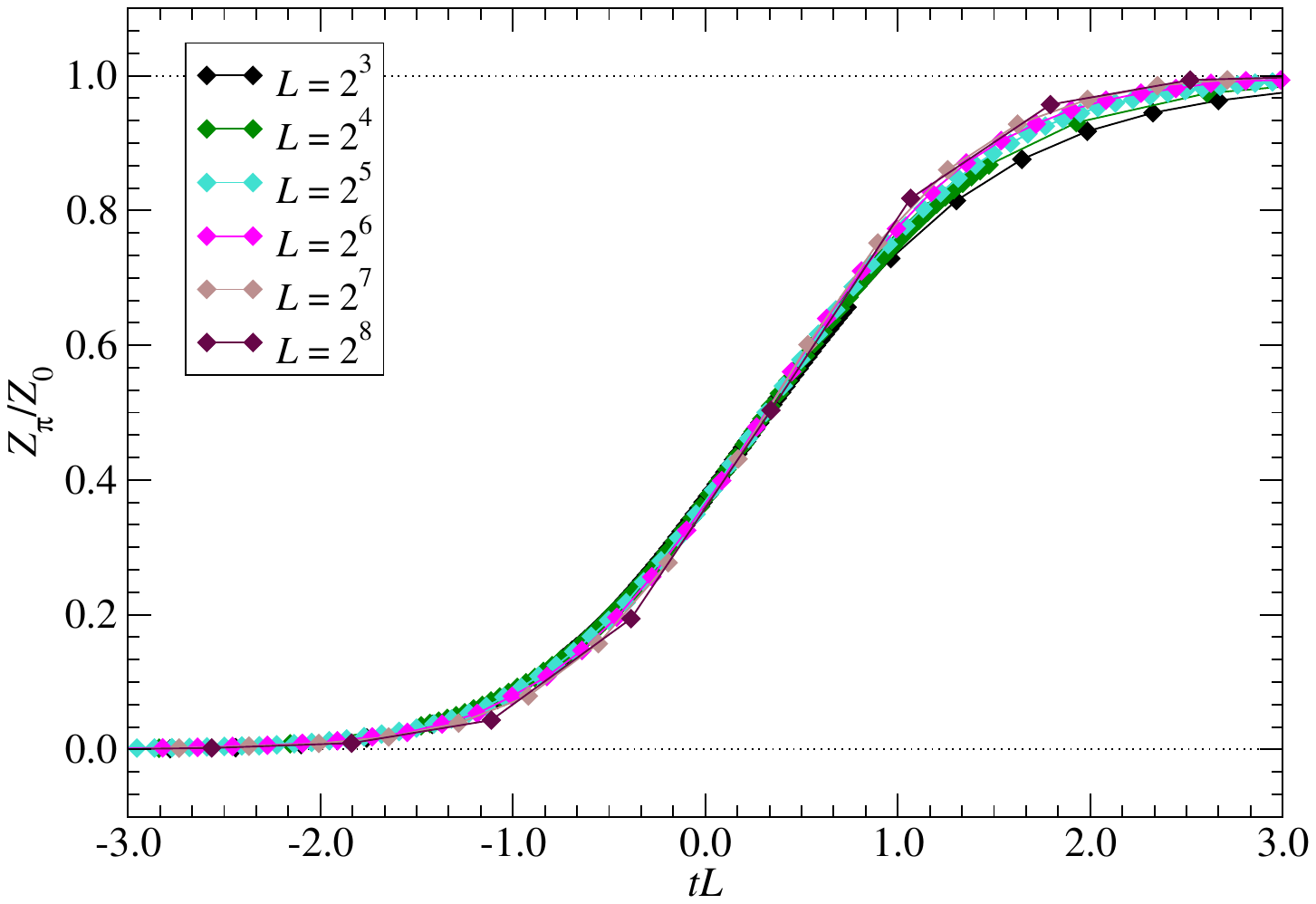}
    }
    \caption{
        BTRG results for the two-dimensional generalized $O(2)$ model ($q=2$, $\Delta=0.16$) with bond dimension $D_{\rm BTRG}=200$.
        (a) $Z_{\pi}/Z_{0}(L,L)$ against temperature, varying $L$.
        The red horizontal line denotes the exact value of $Z_{\pi}/Z_{0}$ for the two-dimensional Ising universality class.
        (b) Finite-size scaling of $Z_{\pi}/Z_{0}$.
    }
    \label{fig:2d_go2_Pi} 
\end{figure}

Fig.~\ref{fig:2d_go2_Pi}(a) shows $Z_{\pi}/Z_{0}$ as a function of temperature for various lattice sizes $L$.
The scale-invariant crossing point of $Z_{\pi}/Z_{0}$ indicates that a second-order phase transition takes place, and the value of $Z_{\pi}/Z_{0}$ at the critical point is consistent with that of the two-dimensional Ising universality class.
Following the finite-size scaling analysis of Ref.~\cite{doi:10.1143/JPSJ.62.435} and assuming the two-dimensional Ising universality class, we perform a finite-size scaling analysis of $Z_{\pi}/Z_{0}$, as shown in Fig.~\ref{fig:2d_go2_Pi}(b), to determine the critical point.
We obtain $T_{c}=0.35253(7)$, which yields a determination with significantly smaller uncertainty than $T_{c}=0.34(1)$ reported in Ref.~\cite{Samlodia:2024kyi}, where the critical temperature was extracted from the analysis of the specific heat.

\begin{figure}
    \centering
    \subfigure[]{
    \includegraphics[width=7.3cm]{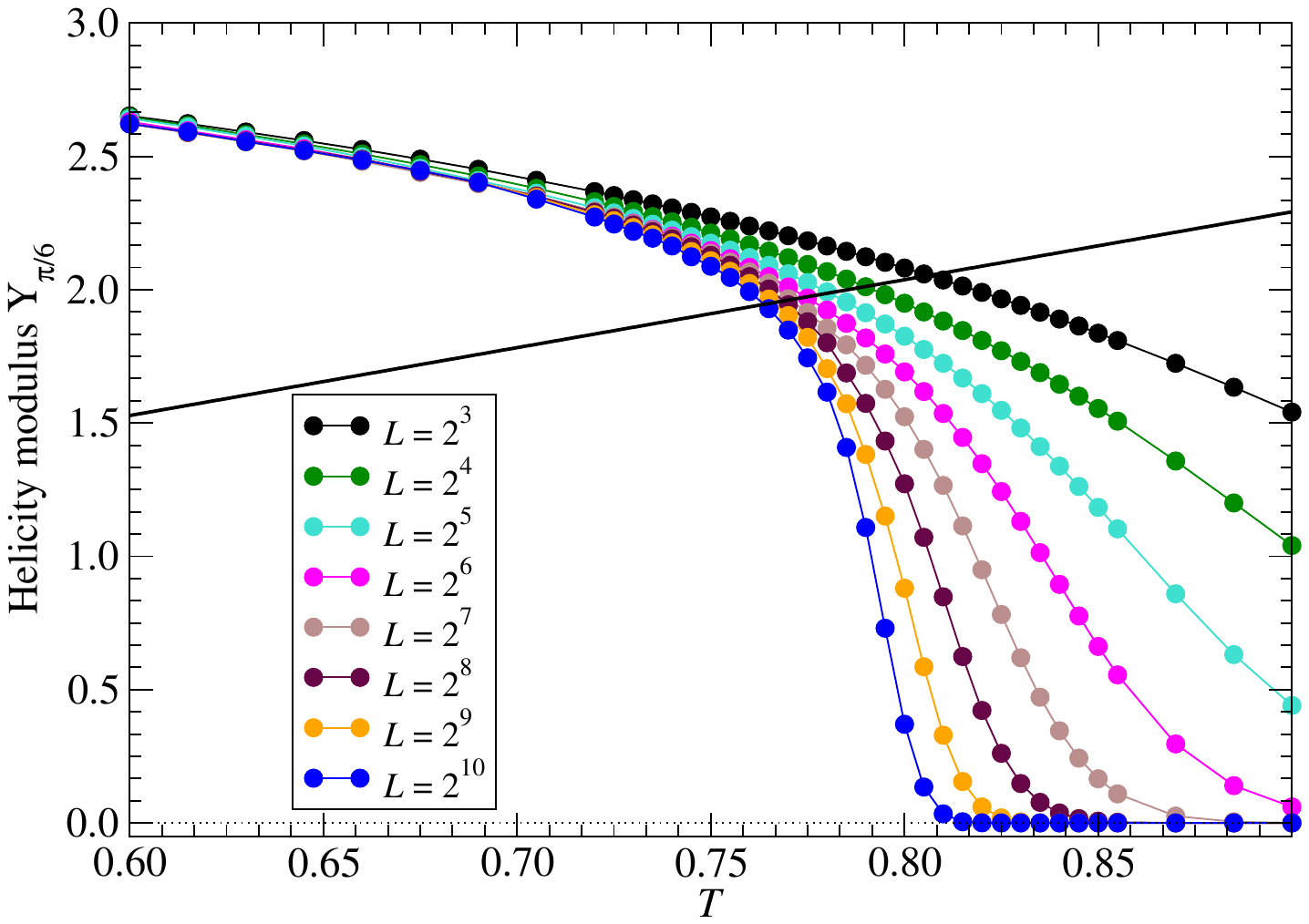}
    }
    \subfigure[]{
    \includegraphics[width=7.3cm]{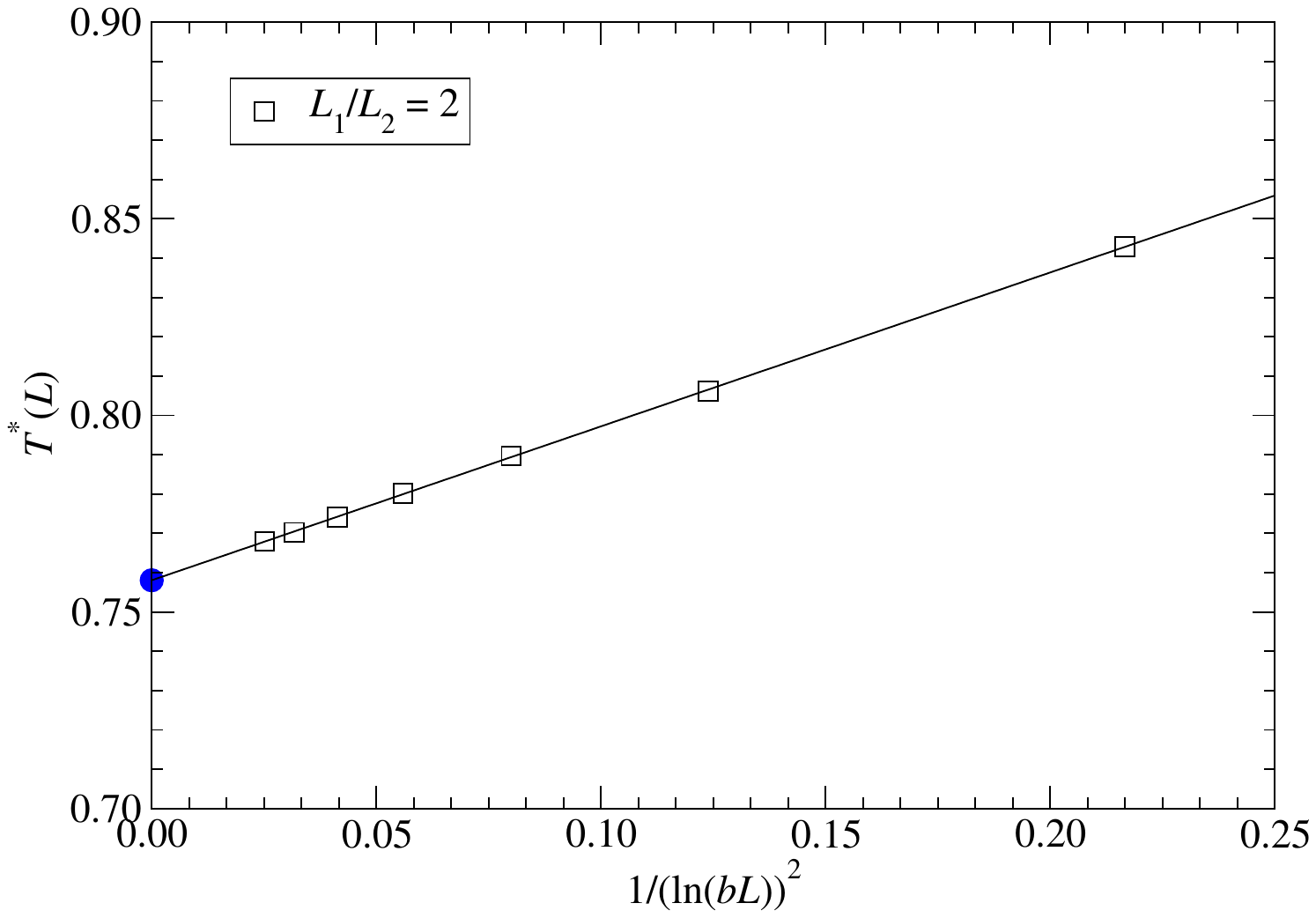}
    }
    \caption{
        BTRG results for the two-dimensional generalized $O(2)$ model ($q=2$, $\Delta=0.16$) with bond dimension $D_{\rm BTRG}=200$.
        (a) $\Upsilon_{\pi/6}(2L,L)$ against temperature, varying $L$.
        The solid line shows the NK criterion, $\Upsilon_{\alpha}=2q^{2}T/\pi$.
        (b) $T^{*}(L)$ as a function of $1/(\ln bL)^{2}$. 
        Squares represent $T^{*}(L)$, and the circle denotes the extrapolated $T_{\rm BKT}$ at $L\to\infty$.
    }
    \label{fig:2d_go2_z12} 
\end{figure}

We also determine the BKT transition at the higher temperature.
Fig.~\ref{fig:2d_go2_z12}(a) shows the helicity modulus $\Upsilon_{\pi/6}$ defined in Eq.~\eqref{eq:Y}.
We have confirmed that $\alpha=\pi/6$ is sufficiently small to suppress finite-$\alpha$ effects in the helicity modulus.
The resulting helicity modulus $\Upsilon_{\pi/6}$ exhibits behavior similar to that of the two-dimensional $O(2)$ model presented in Sec.~\ref{subsec:2d_o2}.
For $q=2$, half-vortices with charge $1/2$ become the relevant topological excitations in the nematic phase.
In this case, the universal jump of the helicity modulus, or equivalently, the NK criterion, is modified from $2T/\pi$ to $2q^{2}T/\pi$.
With this modification, we can follow the same procedure as in Sec.~\ref{subsec:2d_o2} to determine $T_{\rm BKT}$ in the present case as well.
The thermodynamic-limit extrapolation of $T^{*}(L)$ to $T_{\rm BKT}$ is shown in Fig.~\ref{fig:2d_go2_z12}(b).
Our estimate yields $T_{\rm BKT}=0.7581(4)$.
In a previous TRG study, the BKT transition temperature was estimated by introducing an external symmetry-breaking field and analyzing the magnetic susceptibility, yielding $T_{\rm BKT}=0.773(5)$~\cite{Samlodia:2024kyi}.
In contrast to the conventional approach, the symmetry-twisted partition functions allow us to investigate the BKT transition without introducing an explicit symmetry-breaking term.
The absence of an external field reduces the computational cost, making our approach numerically advantageous and also more accurate.

\section{Summary and outlook}
We have demonstrated that the symmetry-twisted partition functions serve as order parameters for identifying phase transitions associated with continuous symmetry breaking, which are difficult to detect using the conventional Gu--Wen ratio. 
As concrete demonstrations, we investigated the three-dimensional $O(2)$ model, where the critical point and the critical exponent $\nu$ were determined through a finite-size scaling analysis of $Z_{\pi}/Z_{0}$. 
We also studied the two-dimensional $O(2)$ model solely using symmetry-twisted partition functions, which directly yield the helicity modulus at a finite twist angle. 
The resulting estimate of $T_{\rm BKT}$ is consistent with previous results in the literature. 
Furthermore, we extended our analysis to the two-dimensional generalized $O(2)$ model with $q=2$. Using only the ratio $Z_{\alpha}/Z_{0}$, we located two phase transition points: one between the ferromagnetic and nematic phases, and the other between the nematic and paramagnetic phases. 
Our results suggest that symmetry-twisted partition functions can indeed identify the phase transitions in the generalized $O(2)$ model by appropriately choosing the twist angle $\alpha\in(2\pi/q)\mathbb{Z}$ or $\alpha\notin(2\pi/q)\mathbb{Z}$.

In future work, we aim to complete the phase diagram of the two-dimensional generalized $O(2)$ model using symmetry-twisted partition functions, not only for $q=2$ but also for $q>4$. In the latter case, unusual ferromagnetic phases are expected, characterized by spins acquiring preferred orientations within a half-plane. 
This implies that the topology of the phase diagram in the $T$-$\Delta$ plane differs from that for $q \le 4$~\cite{PhysRevLett.106.067202}.
In addition, exploring these generalized $O(2)$ models in three dimensions is also an interesting direction and to our knowledge, has not been pursued using TRG methods.

\begin{acknowledgments}
The numerical computation for the present work was carried out with SQUID at the Cybermedia Center, Osaka University (Project ID: hp250055, G16353), and with Pegasus provided by the Multidisciplinary Cooperative Research Program of Center for Computational Sciences, University of Tsukuba.

S. A. acknowledges the support from JSPS KAKENHI Grant Numbers JP23K13096 and JP25H01510, the Center of Innovations for Sustainable Quantum AI (JST Grant Number JPMJPF2221), the Endowed Project for Quantum Software Research and Education, the University of Tokyo~\cite{qsw}, and the Top Runners in Strategy of Transborder Advanced Researches (TRiSTAR) program conducted as the Strategic Professional Development Program for Young Researchers by the MEXT. 
R.~G.~J is supported by the U.S. Department of Energy, Office of Science, Advanced Scientific Computing Research, under contract number DE-SC0025384. 
J. M. is supported by JST SPRING, Grant Number JPMJSP2110.
The work of Y.~T. is partially supported by JSPS KAKENHI Grant No. 23K22489 and also by Center for Gravitational Physics and Quantum Information (CGPQI) at Yukawa Institute for Theoretical Physics (YITP). 
This manuscript has been co-authored by an employee of Fermi Research Alliance, LLC under Contract No. DE-AC02-07CH11359 with the U.S. Department of Energy, Office of Science, Office of High Energy Physics.  This work is supported by the Department of Energy through the Fermilab Theory QuantiSED program in the area of ``Intersections of QIS and Theoretical Particle Physics''.
\end{acknowledgments}

\bibliographystyle{JHEP}
\bibliography{bib/ref}

\end{document}